\documentclass[12pt]{iopart}

\usepackage{graphicx}

\begin{document}

\title[Elliptic flow from pQCD+saturation+hydro model]{Elliptic flow from pQCD+saturation+hydro model}

\author{K J Eskola, H Niemi and P V Ruuskanen}

\address{Department of Physics,  University of Jyv\"askyl\"a, Finland\\
        Helsinki Institute of Physics, University of Helsinki, Finland}


We have previously predicted multiplicities and transverse momentum spectra for the most
central LHC Pb+Pb collisions at $\sqrt{s_{NN}}=5.5$~TeV using pQCD + saturation + hydro (EKRT model)~\cite{Eskola:2005ue, Eskola:1999fc}. We now extend these
calculations for non-central collisions and predict low-$p_{T}$ elliptic flow. Our model is in good agreement with RHIC data
for central collisions, and we show that our extension of the model is also in good agreement 
with minimum bias $v_{2}$ data from RHIC Au+Au collisions at $\sqrt{s_{NN}}=200$~GeV.

We obtain the primary partonic transverse energy production and the formation time in central AA collisions from
the EKRT model~\cite{Eskola:1999fc}.
With the assumption of immediate thermalization we can use these to estimate the initial state for hydrodynamic evolution. 
For centrality dependence we consider here two limits which correspond to models eWN and eBC in~\cite{Kolb:2001qz}, where 
the profile and normalization are obtained from optical Glauber model, once the parameters in central collisions are fixed.
In the eWN (eBC) model the energy density profile and normalization are proportional to the density and the number of wounded nucleons (binary collisions), respectively.
These energy density profiles are used to initialize boost invariant hydro code
with transverse expansion. We use the bag model equation of state with massless gluons and quarks ($N_f =3$), and hadronic phase with all hadronic states
up to a mass $2$ GeV included. Phase transition temperature is fixed to 165 MeV. Decoupling is calculated using standard Cooper-Frye formula, and
all decays of unstable hadronic states are included. Freeze-out temperature is fixed from RHIC
$p_{T}$ spectra for the most central collisions and is $150$~MeV for binary
collision profile~\cite{Eskola:2005ue} and $140$~MeV for wounded nucleon profile. The same freeze-out temperatures are used at the LHC.
Both initializations give a good description of the low-$p_{T}$ spectra for different centralities at RHIC.

\begin{figure}
\hspace{-1.0cm}
\begin{minipage}[b]{0.6\linewidth}
\centering
\hspace*{+0.1cm}
\includegraphics[width=8.2cm]{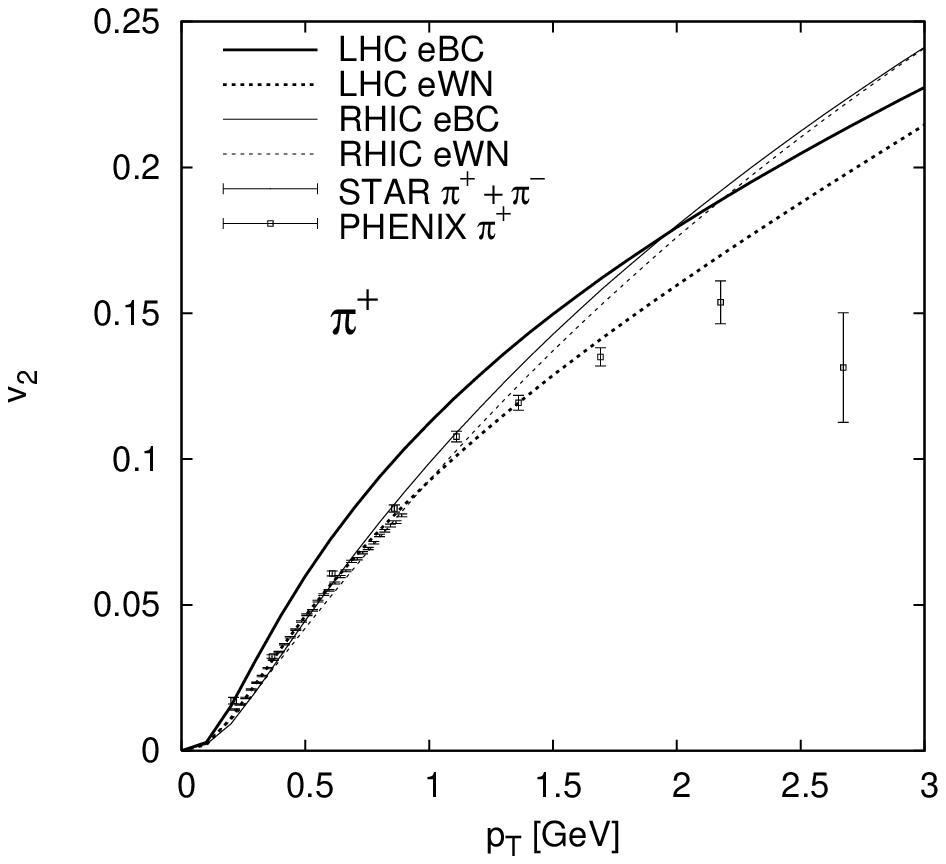}
\end{minipage}
\begin{minipage}[b]{0.6\linewidth}
\centering
\hspace*{-2.8cm}
\includegraphics[width=8.2cm]{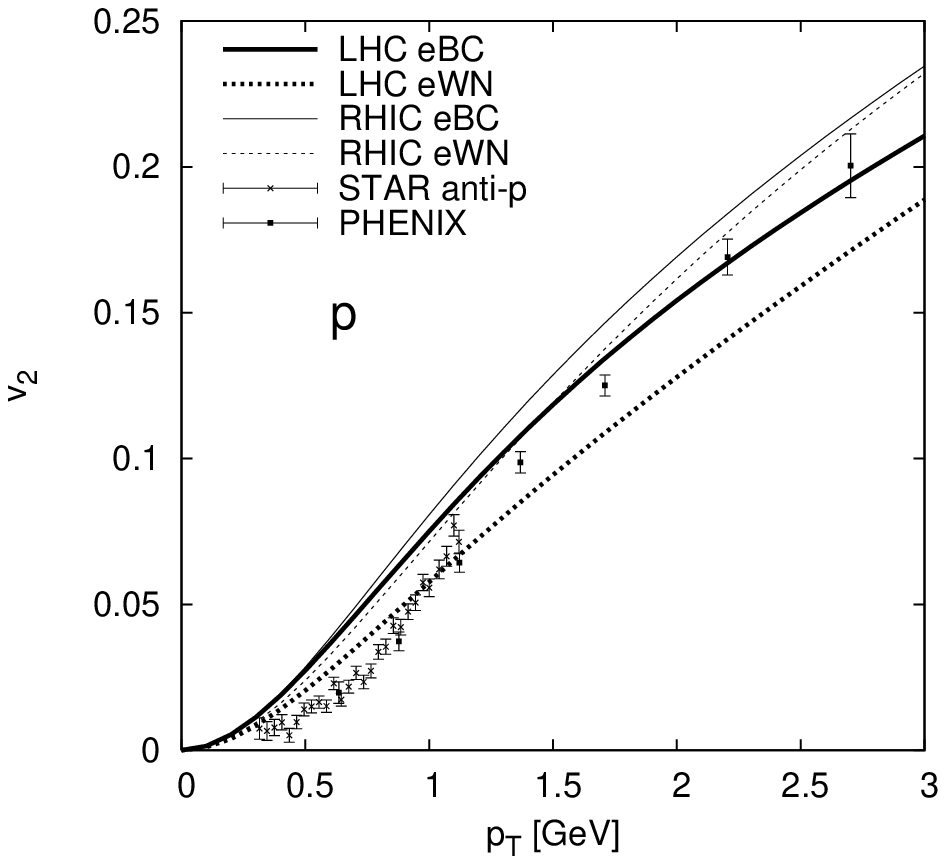}
\end{minipage}
\end{figure}

Left figure shows our calculations for $p_{T}$ dependence of minimum bias $v_{2}$ for positive pions. RHIC results 
are compared with STAR~\cite{Adams:2004bi} and PHENIX~\cite{Adler:2003kt} data. Our minimum bias centrality selection ($0-80$\%) corresponds to
the one used by STAR collaboration. Solid lines are calculations with the eBC model and
dashed lines are from the eWN model. Thin lines are our results for RHIC Au+Au collisions at $\sqrt{s_{NN}}=200$~GeV and thick lines
show our predictions for the LHC Pb+Pb collisions at $\sqrt{s_{NN}}=5.5$~TeV. Largest uncertainty in $v_{2}$ calculations for 
pions comes here from the initial transverse profile of the energy density. Sensitivity to initial time and
freeze-out temperature is much weaker. In general the eWN profile leads to weaker elliptic flow than the eBC profile.
At the LHC the lifetime of the QGP phase is longer, which results in stronger flow asymmetry than at RHIC.
On the other hand the magnitude of transverse flow is also larger, which decreases the $v_{2}$ value at fixed $p_{T}$.
The net effect is that, for a given profile, $v_{2}$ of low-$p_{T}$ pions is larger at the LHC than at RHIC.
Since jet production at the LHC starts to dominate over the hydrodynamic spectra at larger $p_{T}$ than at RHIC~\cite{Eskola:2005ue}, we
expect that the hydrodynamic calculations should cover a larger $p_{T}$ range at the LHC. Thus we predict that the minimum bias $v_{2}$ of pions at fixed $p_{T}$ 
is larger at the LHC than at RHIC, and can reach values as high as $0.2$.

Our model clearly overshoots the proton $v_{2}$ data from STAR~\cite{Adams:2004bi} and PHENIX~\cite{Adler:2003kt}.
A more detailed treatment of the hadron gas dynamics and freeze-out is needed to describe both the proton spectra 
and elliptic flow simultaneously. However, we can still predict the {\it change} in the behaviour of $v_{2}$ of protons when going
from RHIC to the LHC. This is shown in the r.h.s. figure. Although the flow asymmetry increases at the LHC, for more massive particles like protons 
the overall increase in the magnitude of radial flow is more important than for light pions. This results in smaller $v_{2}$ at the LHC than at RHIC
in the whole $p_{T}$ range for protons. Even if $v_{2}$ at fixed $p_{T}$ is smaller at the LHC, $p_{T}$-integrated $v_{2}$ is always 
larger at the LHC for all particles, due to the increase in the relative weight at larger $p_{T}$'s.

\end{document}